\documentclass[3p,time]{./elsarticle}
\usepackage{./ecrc}
\usepackage{rotating}
\usepackage{times}
\usepackage{graphicx}
\usepackage{amssymb}
\usepackage{ulem}
\volume{00}
\firstpage{1}
\journalname{Communications in Nonlinear Science and Numerical Simulation}
\runauth{Portegies Zwart \& Boekholt}

\input journals.def

\def\aplt{\ {\raise-.5ex\hbox{$\buildrel<\over\sim$}}\ }

\def\rm#1{{\sout{#1}}}

\begin{document}
\begin{frontmatter}
  
  \title{
      Numerical verification of
      the microscopic time reversibility of Newton's
  equations of motion: Fighting exponential divergence}
\author{Simon~F.~Portegies~Zwart$^{1}$
        Tjarda~C.~N.~Boekholt$^{2}$}

\address{$^{1}$Leiden Observatory, Leiden University, PO Box 9513, 2300 RA, Leiden, The Netherlands. \\
$^{2}$CIDMA, Departamento de F\' isica, Universidade de Aveiro, Campus de Santiago, 3810-193 Aveiro, Portugal.}

\begin{abstract}

Numerical solutions to Newton’s equations of motion for chaotic self
gravitating systems of more than 2 bodies are often regarded to be
irreversible.  This is due to the exponential growth of errors
introduced by the integration scheme and the numerical round-off in
the least significant figure.  This secular growth of error is
sometimes attributed to the increase in entropy of the system even
though Newton's equations of motion are strictly time reversible.  We
demonstrate that when numerical errors are reduced to below the
physical perturbation and its exponential growth during integration
the microscopic reversibility is retrieved.  Time reversibility itself
is not a guarantee for a definitive solution to the chaotic N-body
problem.  However, time reversible algorithms may be used to find
initial conditions for which perturbed trajectories converge rather
than diverge.  The ability to calculate such a converging pair of
solutions is a striking illustration which shows that it is possible
to compute a definitive solution to a highly unstable problem.  This
works as follows: If you ({\tt i}) use a code which is capable of producing
a definitive solution (and which will therefore handle converging
pairs of solutions correctly), ({\tt ii}) use it to study the statistical
result of some other problem, and then ({\tt iii}) find that some other code
produces a solution {\cal S} with statistical properties which are
indistinguishable from those of the definitive solution, then solution
{\cal S} may be deemed veracious.

\end{abstract}

\end{frontmatter}


\section{Introduction}

General analytic solutions to problems in Newtonian dynamics
\cite{Newton:1687} can only be achieved for a single particle, $N=1$,
or for $N=2$ \citep[][see \cite{doi:10.1119/1.4867608} for a
  historical overview]{Euler1760,Lagrange1772}. Families of periodic
solutions exist for $N>2$ \cite{0951-7715-11-2-011}, and in particular
the parameter-space search of \cite{2017arXiv170904775L} has
succesully identified more than 1000 new periodic solutions to the
restricted 3-body problem, suggesting that the number of such
solutions is interminable.  The latter is consistent with the
existence of periodic solutions within the KAM theorem
\cite{1954Kolmogorov,1962Moser,Arnold1963}.  For all other solutions
approximate methods have to be employed.  These approximate solutions
are unreliable due to the intrinsic chaotic nature of the problem,
leading to exponential growth of small perturbations
\cite{1892mnmc.book.....P,1899lmnd.book.....P,1908BSAFR..22..389P}.
This notion is not new, as Poincare
\cite[][p.\,138]{1903mnm..book.....P} already pointed out:\\
\begin{itemize}
  \item[] {\it Une cause tr\`es petite, qui nous \'echappe,
    d\'etermine un effet consid\'erable que nous ne pouvons pas ne pas
    vois, et alors nous disons que cet effect est d\^u au hasard. Si
    nous connaissions exactement les lois de la nature et la situation
    de l'Universe \`a l'instant initial, nous pourrions pr\'edire
    exactement la situation de ce m\^eme Universe a un instant
    ulterieur. Mais, lors m\^eme que les lois naturelles n'auriaient
    plus de secret pour nous, nous ne pourrions conn\^aitre la
    situation initiale qu'approximativement. Si cela nous perment de
    pr\'evoir la situation ulterieure avec la m\^eme approximation,
    c'est tout ce qu'il nous faut, nous disons que le ph\'enom\`enon a
    \'et\'e pr\'evu, qu'il est r\'egi par des lois; mais il n'en est
    pas toujours ainsi, il peut, arriver que de petites diff\'erences
    dans les conditions initiales en engendrent de tr\`es grandes dans
    les ph\'enom\`enes finaux; une petitt erreur sur les premi\`eres
    produirait une erreur \'enorme sur les derniers. La pr\'ediction
    devient impossible et nous avons le ph\'enom\`ene fortuit.
  }\footnote{ A tiny difference, which we didn't notice, has
    considerable repercussions that we can't ignore, and then we say
    that this effect is pure chance.  If we knew exactly the laws
    of nature and the conditions of the universe at the initial
    moment, we could exactly predict the situation of that same
    universe at any succeeding moment. But even if it were the case
    that the natural laws had no longer any secret for us, we could
    still only know the initial situation approximately. If that
    enabled us to predict the succeeding situation with the same
    approximation, that is all we require, and we should say that the
    phenomenon had been predicted, that it is governed by laws. But it
    is not always so; it may happen that small differences in the
    initial conditions produce very great ones in the final
    phenomena. A small error in the former will produce an enormous
    error in the latter. Prediction becomes impossible, and we have
    the fortuitous phenomenon.}
\end{itemize}

\noindent
As a result, small errors in the temporal or spatial discretizations,
or in the numerical integration scheme cause any solution to
eventually become invalid \cite{1964ApJ...140..250M}, maybe even
within a few dynamical time scales \cite{1986LNP...267..212D}.

The relation between instability and chaoticity manifests itself by
the high sensitivity to small changes in the initial conditions
\cite{Hadamard1898,1908BSAFR..22..389P,1963JAtS...20..130L}.
Divergent behavior is often demonstrated by performing two
calculations with a slight offset $\delta$ in phase space.  The
exponential growth with time of this phase-space distance $\delta(t)$
is expressed in the largest Lyapunov exponent \cite{lyapunov1892}.
For a chaotic system, the associated e-folding time is positive and
finite.  The rate of divergence of two trajectories in phase space is
characterised by this $e$-folding time, which is specific for the
$N$-body realization.  Since Newton's equations of motion are time
reversible a finite parameter space must exist for which $\delta(t)$
diminishes \cite{1930Perron}; namely, when an initial divergent pair
of trajectories is time reversed, it should return to the moment the
perturbation was introduced. Such behavior can only be established
when solving Newton's equations of motion with sufficient accuracy and
enough precision to guarantee that the accumulated error and its
exponential growth remains below the introduced deviation
$\delta(t=0)$.  In Appendix A we present a short
glossary of terms used in the manuscript to help the reader appreciate
our disucssion.

The combination of hypersensitivity to small perturbations and
non-integrability prevents us from manifesting Newton's reversibility
numerically, because the underlying method should be accurate as well
as precise in order to arrive at a converged solution.  High-order and
symplectic numerical solvers tend to be insufficiently accurate, in
the sense that reducing the time step tends to interfere with the lack
of precision due to the growth of the error introduced by round off
\cite{2015ComAC...2....2B}.  The combination of the exponential growth
of small perturbations and the inevitability of numerical errors forms
the fundamental argument why solving Newton's equations of motion
still is one of the hardest problems in computational physics.
Individual numerical trajectories quickly forget their initial
conditions\footnote{With the term ``forget'' we mean that the
  behaviour of a solution (in some chaotic region) becomes
  statistically independent of its initial conditions (see
  \cite{2014ApJ...785L...3P}).}.  Instead of integrating a single
realization until a converged solution is obtained one often considers
ensembles of trajectories in phase space, starting with a random
sample of initial realizations close to, and possibly including, the
objected realization \cite{Szebehely1991}.  Each of these realizations
is subsequently calculated and the objected phase space is anticipated
to provide a probability density distribution around the true
solution.  In principle this leads to a veracious solution, but this
is not guaranteed, in which case the ensamble average may well be
distinct from the true solution.  \cite{2015ComAC...2....2B}
demonstrated that an ensamble of reprehansible solutions was
statistically indistringhuisable from the an ensable of converged
solutions with idential realizations, a quality we call {\it nagh
  Hoch}.

It is not clear whether the strict time-reversibility of Newton's
equations of motion can be feasibly maintained from a practical
numerical point of view.  The authors in
\cite{2008MNRAS.388..965L,2016ttp..book.....V}, claim that the chaotic
nature of the underlying dynamical processes then prevents us, via the
second law of thermodynamics, to reverse time and calculate backward.
\cite{2008MNRAS.388..965L} associate such irreversible dynamical
process to the increase of entropy and the arrow of time in systems
that, from a theoretical perspective should be strictly
deterministic. We instead call this process numerical confusion.

Chaos in $N$-body systems is often confused with a number of side
effects, including uncertainties in the initial conditions, round-off,
integration errors, spatial and temporal discretization, as was
already pointed out by \cite{1991pscn.proc...47H}.  In a chaotic
self-gravitating system errors grow exponentially but with sufficient
accuracy and precision this does not prevent the system from resulting
in a definitive solution.  Such calculations are time reversible
\cite{1994CeMDA..60..365A}, but time reversibility itself is
insufficient to guarantee a definitive result because the method may
still fail to resolve close encounters either by lack of accuracy or
by lack of precision.  Two neighboring solutions on the other hand
recover their initial offset when reversed after some finite time.  In
such a time-reversed evolution trajectories approach each other until
the initially introduced phases-space distance $\delta(t=0)$ is
reached, after which they diverge again.  Such a phase-space offset
deminishing simulation leads to exponential convergence within a
finite time interval.  Most studies in $N$-body dynamics focus on the
largest positive Lyapunov exponent to measure the rate of divergence
between neighbouring trajectories rather than on converging solutions
\citep{LichtenbergAndLieberman1992}.  In the next \S\, we demonstrate
that such converging trajectories exist, and we speculate that they
are important when the duration of the convergence phase becomes
comparable to the simulation time-scale of interest.

\section{Results for the Pythagorean 3-body problem}

\begin{figure}
\begin{center}
\includegraphics[width=1.0\textwidth]{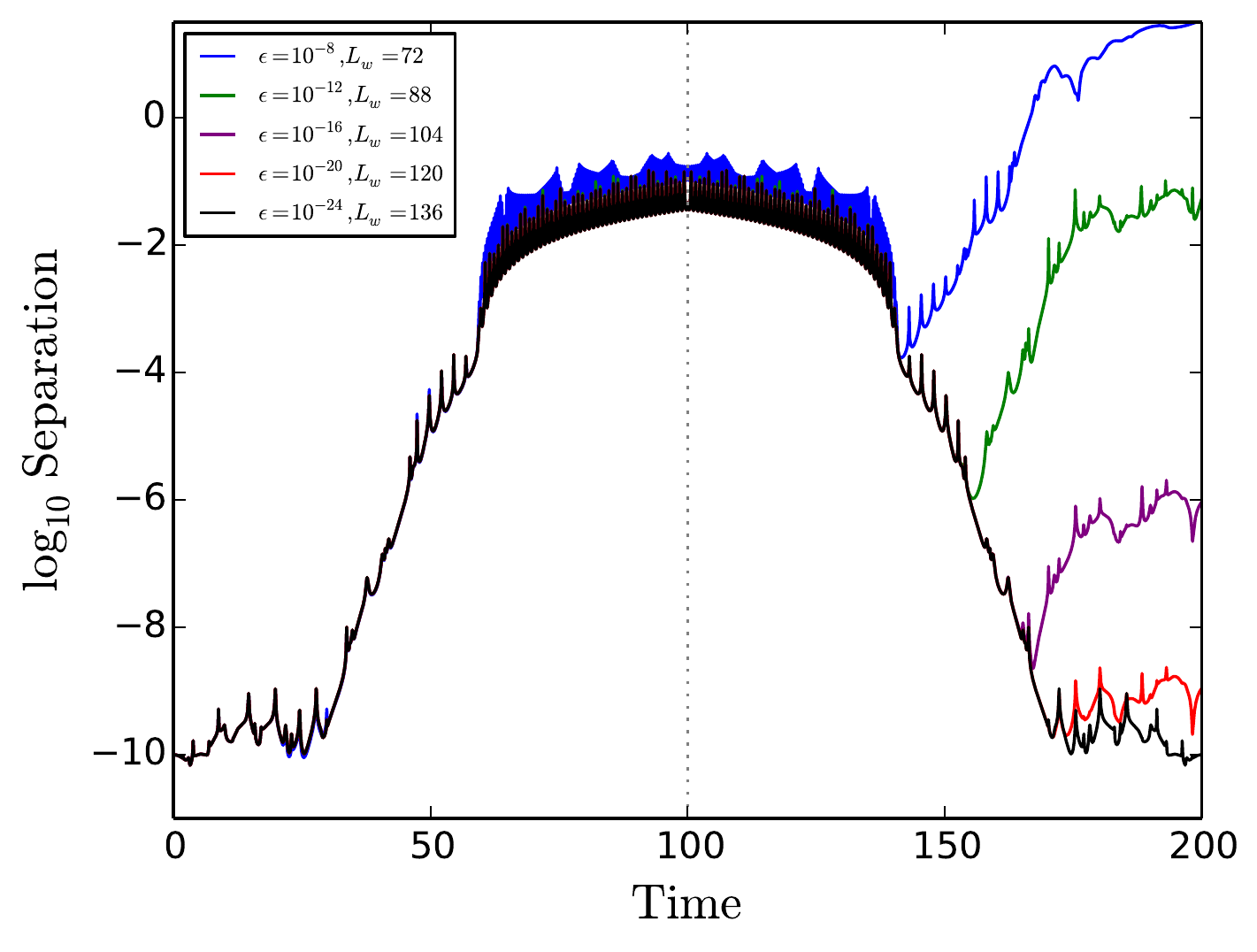}
\caption{Evolution of the phase-space separation $\delta$ between two
  solutions of the Pythagorean $3$-body problem
  \cite{1913AN....195..113B}.  The perturbed trajectories are
  calculated using an initial offset of $10^{-10}$ in the x-coordinate
  of the least massive body (see Tab.\,\ref{ICs}).  After $t=100$ we
  reverse all the velocities and continue the integration to $t=200$.
  The accuracy of the Bulirsch-Stoer integrator is tuned using two
  parameters: the tolerance parameter $\epsilon$ and the word-length
  $\eta$.  The identical evolution of both systems back to their
  initial conditions is achieved using a Bulirsch-Stoer tolerance of
  $10^{-24}$ and a word-length of 136\,bits (see Tab.\,\ref{Final} for
  the interim and final results).  Less accurate and less precise
  calculations are presented with the colored curves (see top left
  inset for the parameters).
\label{Fig:errorproagation} 
}
\end{center}
\end{figure}

Numerical verification of the time reversibility of Newton's equations
of motion can be obtained by demonstrating the existence of converging
trajectories in phase space, and the ability to calculate any
non-colliding initial condition to an arbitrary point in time and
backward to recover the initial realization with an accumulated
integration error much smaller than the initial offset.  In colliding
orbits, such as the case in homothetic hyperbolic orbits
\cite{10.2307/24893288,1974InMat..27..191M}, the singularity can only
be reached asymptotically which restricts the time-scale over which
the solution can be obtained numerically.  In
Fig.\,\ref{Fig:errorproagation} we demonstrate numerical
time-reversibility by presenting the result for the Pythagorean
$3$-body problem \cite{1913AN....195..113B}, which was first solved in
1967 \cite{1967AJ.....72..876S} and which is chaotic
\cite{1994CeMDA..58....1A} but relatively short lived. In
Tab.\,\ref{ICs} we present an initial realization for the Pythagorean
problem (and the perturbed conditions). In the Appendix B we present a
brief summary of the results of the calculations. Snapshots of the
converged solution are available in the online supplementary material.
The possibility to time-revers the calculations apply to any
self-gravitating multi-body system and the choice of the Pythagorean
problem is motivated by limited resources and its historical context.
We integrate the equations of motion for the $3$-bodies using the {\tt
  Brutus} arbitrary precise $N$-body code
\cite{2015ComAC...2....2B,2016MNRAS.461.3576B}, which is part of the
AMUSE software environment
\cite{2013CoPhC.183..456P,2013A&A...557A..84P}.  In {\tt Brutus} the
integration accuracy is controlled by the tolerance $\epsilon$ in the
Bulirsch-Stoer \cite{springerlink:10.1007/BF01386092} integrator, and
the precision is controlled using length of the mantissa $\eta$.  By
changing these, {\tt Brutus} can be tuned until the solution
converges. Finding a converged solution is an iterative procedure in
which the values of the Bulirsch-Stoer tolerance $\epsilon$ and the
length of the mantissa $\eta$ are improved until to a pre-determined
precision the phase-space coordinates of the solution becomes
independent of these parameters.

\begin{table}[]
\centering
\caption{Initial realization of the Pythagorean problem as adopted in
  our numerical experiment.  Each line contains mass, $x$- and
  $y$-position and $x$- and $y$-velocity in units in which we adopt
  $G=1$. The $z$-coorindates are zero and omitted from the table.  The
  mit-time and final simulation results are presented in the Appendix
  B.  }
\label{ICs}
\begin{tabular}{l|ll|ll}
  \hline
  \hline
 & \multicolumn{4}{l}{Unperturbed initial conditions}\\  
  \hline
  mass & \multicolumn{2}{c}{position ($x$, $y$)} & \multicolumn{2}{c}{velocity ($v_x$, $v_y$)}  \\
  3 & 1 & 3  &0 &0  \\
  4 & -2& -1 &0& 0 \\
  5 & 1 &-1  &0 &0 \\
  \hline
 & \multicolumn{4}{l}{Perturbed initial conditions}\\  
  \hline
  3 & 1.0000000001 & 3 &0 &0   \\
  4 & -2& -1 &0& 0 \\
  5 & 1 &-1  &0 &0  \\
  \hline
\hline
\end{tabular}
\end{table}

We integrate the Pythagorean problem for $100$ time-units ($G=1$, the
units of mass and length are implicitly defined in Table\,\ref{ICs}):
any time and any initial realization would have worked but may require
a different tolerance and word-length in the integration scheme
\cite{2015ComAC...2....2B}. Two trajectories are calculated, for one
we offset the $x$-position of the least massive body by $\delta =
10^{-10}$ (see Tab.\,\ref{ICs}) but a similarly small offset in any of
the Cartesian coordinates would have worked. The three bodies engage
in a resonant interaction, which lasts until one particle
escapes\footnote{A particle is considered to escape when it is
  receding and its binding energy with respect to the system is
  negative \cite[see][]{1983ApJ...268..319H}.} at $t \simeq
63.4^{\ref{Footnote:Dissolution}}$, consistent with earlier results
\cite{1967AJ.....72..876S,1986LNP...267..212D,1994CeMDA..58....1A,2008MNRAS.386..425F}.
During this period two distinct chaotic behaviors manifest themselves.
The system starts with a relatively long $e$-folding time scale of
$\sim 28.9$ of the exponential growth of the initial phase-space
separation (see Fig.\,\ref{Fig:errorproagation}).  After 28.6 time
units a transition occurs, possibly initiated by a close three-body
encounter (see also \citep{2008MNRAS.386..425F}), leading to a faster
growth of the phase-space distance. In this lap the system exhibits a
much shorter $e$-folding time scale of $\sim 2.1$. The unperturbed and
the perturbed systems eventually dissolve into a bound pair and a
single body\footnote{The dissolution time was determined by fitting an
  exponential plus a linear model to $\delta$ as a function of
  time. At $t\simeq 63.4$ the linear behavior overtakes the
  exponential behavior of $\delta$. The corresponding time scale then
  indicates when the system's diverging characteristics are no longer
  exponential.  For other escape criteria the dissolution time may
  vary.\label{Footnote:Dissolution}}. After this the phase-space
separation grows approximately linearly.  This evolution is consistent
with results of previous studies
\cite{1986LNP...267..212D,1994CeMDA..58....1A}.

We stop the calculation at $t = 100$ by which time the phase-space
separation between the two solutions has grown by a factor of $10^{9}$
to $\delta \sim 0.1$. We subsequently reverse the velocities of all
particles and continue the calculation to $t=200$.  At this moment, we
compare the resulting positions and velocities with the original
initial conditions (see Tab.\,\ref{Final}).  In order to realize a
reversible calculation in which the initially introduced perturbation
grows by 9 orders of magnitude the numerical round-off error and its
exponential growth has to remain well within a factor of $10^{18}$ of
the initially introduced offset.  With a Bulirsch-Stoer tolerance of
at most $10^{-24}$ and a word-length of at least 136\,bits, the final
and initial conditions are identical to a sufficiently large mantissa
to recover the initial offset between the two initial realizations.
In Tab.\,\ref{Final} we present the final conditions for the
unperturbed and the perturbed initial realizations.

\section{Discussion}

With the calculations of the Pythagorean 3-body problem, we
demonstrate that irrespective of the exponential growth of
perturbations the equations of motion can be solved accurately
although not exactly using floating-point arithmetic. Newton's laws of
motion are time reversible and chaos does not prevent this.
Microscopic irreversibility (microscopic in the sense of detailed
positions and velocities), as found in previous studies
\cite{hollinger2012nature,2008MNRAS.388..965L} is then the result of
the secular growth of numerical errors.  From a theoretical
perspective, \cite{MR103254,MR103256} already argued that
irreversibility is not an intrinsic quality of chaos.  Earlier claims
of reversibility in $N$-body simulations \citep{2017arXiv170407715R}
using high-order direct integrators with double floating point
precision results in an evolution of the phase-space distance
comparable to the blue curve in Fig.\,\ref{Fig:errorproagation}, but
much higher precision is needed in order to recover the initial
conditions (black curve).

In $N$-body integrations the numerical error accumulates due to
round-off and discretization errors.  Irreversibility of these
numerical errors will render the entire calculation
irreversible. Integer arithmetic, as alternative to floating points,
could solve this problem because round-off in that case is
deterministic and time symmetric \cite{1992PhyD...56....1E}. This
makes the calculations strict time symmetric, but they remain
reprehansible in the sense that they are not necessarily accurate and
not precise.  We test this hypothesis by repeating the calculations
(see Tab.\,\ref{ICs}) using the {\tt Janus} integrator in the {\tt
  Rebound} package \cite{2017arXiv170407715R}, which employs integer
arithmetic.  We verify that integer arithmetic in the $N$-body problem
leads to a time reversible result, but it is insufficiently accurate
to resolve close encounters, irrespective of the time symmetry in the
round off.  This is evidenced by a growth of the phase-space distance
that deviates from the curve presented in
Fig.\,\ref{Fig:errorproagation}.  In our experiment we explored a time
step size of 0.1 to $10^{-26}$ but the solution did not converge.  We
tested the effect of floating-point behavior on the growth of the
phase-space distance using different integrators, 4th order
predictor-corrector Hermite \citep{1992PASJ...44..141M}, using the
implementation of {\tt ph4}, which is part of the {\tt AMUSE} package
\cite{AMUSE} and the Verlet leap-frog integrator \cite{PhysRev.159.98}
which is part of the {\tt Starlab} package \cite{1999A&A...348..117P}
and found identical behavior up to the point the curves start to
deviate some time after the time-reversed calculation (see the blue
curve in fig.\,\ref{Fig:errorproagation}).

The parameter space for converging trajectories is small.  In our
experiment, where we follow the growth of the perturbation over $9$
orders of magnitude in the normalized $6$-dimensional phase space,
convergence is not reached before we extend the mantissa of the
calculation to exceed 18 decimal places (for calculating forwards in
time and backwards). In order to continue hiding the round-off, the
entire calculation had to be performed with 40 significant figures
(see Tab.\,\ref{Final}).  A naive estimate of the available parameter
space for finding such converging solutions is then $\propto
1/10^{18}$.  The arrow of time then manifests itself by the larger
parameter space for which irreversible perturbations lead to a
divergent growth rather than convergence.

Once we are able to time-reverse a relatively simple three-body
simulation, it remains unclear to what extend this can be attributed
to a complicated multi-body simulation. In a gedanken experiment one
could evolve a star cluster to the moment just before the first hard
binary forms in the collapsing core.  When starting from this point we
would calculate backwards in time, the system should return to its
initial realization. When at this point we introduce a small
perturbation to one of the single objects the system will not return
to its initial realization. But we wonder how large a perturbation are
we allowed to introduce for the system to still climb out of core
collapse, rather than continue to collapse until the actual formation
of the first binary?

\section*{Acknowledgments}

We thank Fouvry Jean-Baptiste, Vincent Icke, Junichiro Makino, Cole
Miller, Inti Pelupessy, Dagmar Portegies Zwart, Hanno Rein, Mauri
Valtonen and the accurate and precise non-anonymous referee for
discussions.  This work was supported by the NWO (grant \#621.016.701
[LGM-II]) the European Union’s Horizon 2020 research and innovation
programme under grant agreement No 671564 (COMPAT project), and
Fundaç\~ ao para a Ci\^ encia e a Tecnologia (grant
SFRH/BPD/122325/2016), and the Center for Research \& Development in
Mathematics and Applications (CIDMA) (strategic project
UID/MAT/04106/2013) and from ENGAGE SKA, POCI-01-0145-FEDER-022217,
funded by COMPETE 2020 and FCT, Portugal.

\vfill
\newpage

\section*{Appendix A}\label{AppendixA}

Here we present a small glossary of terms used in this manuscript.

\begin{itemize}
  \item[] {\bf Accuracy, accurate:} degree to which a calculation is
    numerically accurate in terms of the discretization of the
    numerical method.  Accuracy is controlled by the time step size of
    the integration. In the Bulirsch-Stoer method the accuracy is of
    the order the tolerance $\epsilon$.
    
    \item[] {\bf Converged:} A numerical solution that to a certain
      pre-determined length of the mantissa will not change
      irrespective of the further increase in the mantissa during the
      calculation (precision) or a further decrease of the time step
      (accuracy). The term should be used in conjuction with the
      length of the matissa of the converged simulation.  The
      iterative process that leads to a converged solution is
      considered {\it converging}, which should not be confused with
      the term {\it convergence}.  Converged solutions are considered
      {\it definitive}.

  \item[] {\bf Convergence:} The gradual decrease in the phase-space
    distance between two different solutions over a course of time.

  \item[] {\bf Definitive solution:} The solution for some initial
    realization that is expected to be numerically indistinghuishable
    from the true solution.  From a mathematical perspective a
    definitive solution represents a computable non-periodic
    pseudo-orbit \cite{1987PhLA..122..399P}.
    
\item[] {\bf Initial conditions:} statistical description of boundary
  values from which, using some random number sequence with a seed
  value, an {\it initial realization} can be generated.
           
\item[] {\bf Initial realization:} actual specific phase-space
  coordinates that uniquely describes a system of particles.  Any
  initial realization results after some time in one unique final
  realization.
           
\item[] {\bf nagh Hoch:} The concept that an ensemble of random
  initial realizations in a wide range of parameters gives
  statistically the same result as the converged solutions of the same
  ensemble realizations. This concept is a quality of the numerical
  method, but we speculate that this quality also applies to real
  systems.  nagh Hoch means as much as ``similar appearance''
  \cite{KlingonDictionary}.

  \item[] {\bf neighboring solutions:} Two initial realizations that
    started with an infinitesimal offset integrated over a time scale
    sufficiently short that the exponential growth of the offset
    remains below a predetermined limit (typically small compared to
    the initial size of the system but possibly large compared to the
    initial offset) \cite{1992PhLA..166..340F}.

  \item[] {\bf Precision, precise:} degree to which a calculation is
    numerically reproducible in terms of the length of the
    mantissa. Precision in Brutus is controlled using length of the
    mantissa $\eta$.
    
\item[] {\bf Reprehansible reprehensive:} Solution to Newton's
  equations of motion for which the accumulation of numerical errors
  and the system's response exceeds the exponential growth of the
  initial offset $\delta$.

  \item[] {\bf Time reversibility:} The ability of a numerical
    integrator to recover the initial realization from reversing the
    final realization.  Time reversibility does not guarantee that the
    solution is converged as it can already be obtained by insisting
    time symmetry in numerical errors without resolving close
    encounters.

  \item[] {\bf True solution:} The unique solution Nature would
    provide in the limited physical domain (such as Newtonion dynamics
    in isolation) for a specific initial realization.

  \item[] {\bf Veracity, veracious:} the concept that a limited and
    bounded variation of the initial conditions near a pre-selected
    initial realization gives statistically an indistinguishable
    ensemble average as a single {\it converged} solution.  A
    veracious solution can be perceived as the experimental
    observation of the true solution \cite{2003PhRvL..90e4104H}.  The
    presumption that Newtonian simulations are {\it veracious} is used
    for example in stability studies of the Solar System \cite[see
      e.g.]{2008DDA....39.1514V,2015ApJ...798....8Z}.

\end{itemize}
          
\section*{Appendix B}\label{AppendixB}

The Pythagorean $3$-body problem \cite{1913AN....195..113B} starts
with three point masses of 3, 4, and 5 in the $x$-$y$--plane and with
zero velocities. The three masses are located in the corners of a
Pythagorean triangle with (1, 3) for the first mass, (-2, 1) for the
second and (1, -1) for the third and most massive object.  The values
are dimensionless modified $N$-body units \cite{1986LNP...267..233H}.
In Tab.\,\ref{ICs} we present these coordinates, and the initial
realization of our perturbed solution.

In Fig.\,\ref{Fig:errorproagation} we present the phase-space separation
between the converged solution and a perturbed solution, and in
addition a number of less accurate calculations. The less accurate
solutions do not recover the initial phase-space distance of
$10^{-10}$ but continue to diverge at some moment during the
integration. The moment they start to diverge depends on the adopted
accuracy.

In Tab.\,\ref{Final} we present the mid-point (at $t=100$) and final
conditions (at $t=200$) for the converged unperturbed configuration
and the perturbed initial realization. Here we omitted the third
coordinate ($z$) because it is $0$ and remains $0$ throughout the
calculation. We did not round the numbers but present them as they are
represented in the computer in 40 significant figures.

\begin{sidewaystable}
\centering
\caption{Cartesian coordinates up to the last significant figure as
  represented in the computer (40 decimal places) of our calculation
  at mid-point and the final conditions for the Pythagorean problem.
  The initial conditions are presented in Tab.\,\ref{ICs}.  Each
  particle is represented by 2 lines: the first line includes the
  particle's mass, $x$-position and $x$-velocity, and on the next line
  contains the $y$-position and $y$-velocity.  }
\label{Final}
\begin{tabular}{llll}
\hline
\hline
 mass & position ($x$, $y$) & velocity ($v_x$, $v_y$) \\
\hline
\hline
\multicolumn{3}{l}{Unperturbed configurations}\\
\multicolumn{2}{l}{Midpoint (t=100)}\\
\hline
3& 23.177734271287932984376719864262469663010   &  0.5319181795481124436207125296002406419911\\
 &   68.527351147839055139293277996603512381330 &  1.5806796434841945398880662764620232011290\\
4& -7.259955925491511142826576066493924610372   & -1.3069988754994211535542088320251159535050\\
 &  -22.988301724483254153623924485962763879080 & -0.3946057714182357399709966348235988763510\\
5& -8.098675822379550876364771064194141771170   &  0.7264481926706694566709395478765597018697\\
 & -22.725769309116829760676827209503214745270  & -0.6327231689559281319560424580219478896762\\
\hline
\multicolumn{2}{l}{End (t=200)}\\
\hline
3&	  1.000000000000029147791398925636660776567& 	-2.270447738062808857494824732048534406358e-14\\
 &	  3.000000000000036383772882843757984240521&  	 4.774666017242739047666501771157480010792e-14 \\
4&	 -1.999999999999974915829187984060851093363& 	 4.774665242050519395810021558737175983173e-14\\
 &	 -0.999999999999810155587671818578010637774&  	 2.270447738062669364553550881039426305535e-14 \\
5& 	  0.999999999999962443988511031800475014671& 	-2.457463550802730796268351737568898744656e-14\\
 &	 -1.000000000000173705793592251440952990686&  	-4.681157800795778807649659958570352078546e-14 \\
\hline
\hline
\multicolumn{3}{l}{Perturbed configurations}\\
\multicolumn{2}{l}{Midpoint (t=100)}\\
\hline
3&	 23.190521482793319989765681634995570288100 &	 0.5322503567960587903769290142249037912735\\
 &        68.556547837283584752026354886258680715190& 	 1.5814556205092701449052881349866961910410\\ 
4& 	 -7.287343535103771287105271550570713456589 &	-1.4321791210002474579596089005290112193940\\ 
 &       -22.995007467016918242398592448181673127520&  	-0.3552190002153252430432594844029797100956 \\ 
5& 	 -8.084438061532974964175191740749603394790 &	 0.8263930827225626921415297118904104229181\\ 
 &       -22.737922728756616257296938973228645901360&  	-0.6646981721333018925085652934724455130194 \\ 
\hline
\multicolumn{2}{l}{End (t=200)}\\
\hline
3& 	  1.000000000100049154879867130170780709564 &	-3.828886532300641220319197711163974610107e-14\\
 &         3.000000000000061357649755490016408020941&  	 8.052003949335830235349211152837151414787e-14 \\
4&	 -1.999999999999957698016272046329507322690 &	 8.052002826247233760595438150140762918720e-14\\
 &        -0.999999999999679846094415534203355381441&  	 3.828886532501206909565529941811723471334e-14 \\
5& 	  0.999999999999936665485097358523258172108 &    -4.144270341617403501659375346577704899564e-14 \\
 &        -1.000000000000292937714320866161883344304&  	-7.894311595602463066508120495426148312190e-14 \\
\hline
\hline
\end{tabular}
\end{sidewaystable}


\begin{thebibliography}{10}

\bibitem{Newton:1687}
I.~Newton, {\em Philosophiae Naturalis Principia Mathematica}, vol.~1 (1687).

\bibitem{Euler1760}
L.~{Euler}, {\em Nov. Comm. Acad. Imp. Petropolitanae} {\bf 10}, 207 (1760).

\bibitem{Lagrange1772}
J.-L. {de Lagrange}, `{Chapitre II: Essai sur le Probl\`eme des Trois Corps}'
  (1772).

\bibitem{doi:10.1119/1.4867608}
M.~\v{S}uvakov, V.~Dmitra\v{s}inovi\'c, {\em American Journal of Physics} {\bf
  82}(6), 609 (2014).

\bibitem{0951-7715-11-2-011}
R.~Montgomery, {\em Nonlinearity} {\bf 11}(2), 363 (1998).

\bibitem{2017arXiv170904775L}
X.~{Li}, Y.~{Jing}, S.~{Liao}, {\em ArXiv e-prints}  (2017).

\bibitem{1954Kolmogorov}
A.~N. {Kolmogorov}, {\em Dolk Akad.\, Nauk SSSR} {\bf 98}, 527 (1954).

\bibitem{1962Moser}
J.~{Moser}, pp. 1--20 (1962).

\bibitem{Arnold1963}
I.~{Arnold}, {\em Uspehi Mat. Nauk} p.~18 (1963).

\bibitem{1892mnmc.book.....P}
H.~{Poincare}, {\em {Les methodes nouvelles de la m\'ecanique c\'eleste}}
  (1892).

\bibitem{1899lmnd.book.....P}
H.~{Poincare}, {\em {Les methodes nouvelles de la m\'ecanique c\'eleste}}
  (1899).

\bibitem{1908BSAFR..22..389P}
{Poincar{\'e}}, {\em Bulletin de la Societe Astronomique de France et Revue
  Mensuelle d'Astronomie, de Meteorologie et de Physique du Globe} {\bf 22},
  389 (1908).

\bibitem{1903mnm..book.....P}
H.~{Poincar{\'e}}, {\em {La Science et l'Hypoth\'ese}}.

\bibitem{1964ApJ...140..250M}
R.~H. {Miller}, {\em \apj} {\bf 140}, 250 (1964).

\bibitem{1986LNP...267..212D}
H.~{Dejonghe}, P.~{Hut}, in P.~{Hut}, S.~L.~W. {McMillan} (eds.), {\em The Use
  of Supercomputers in Stellar Dynamics}, vol. 267 of {\em Lecture Notes in
  Physics, Berlin Springer Verlag}, p. 212 (1986).

\bibitem{Hadamard1898}
Hadamard, {\em Journal de Mathématiques Pures et Appliquées} {\bf 4}, 27
  (1898).

\bibitem{1963JAtS...20..130L}
E.~N. {Lorenz}, {\em Journal of Atmospheric Sciences} {\bf 20}, 130 (1963).

\bibitem{lyapunov1892}
A.~Lyapunov, {\em Commun. Math. Soc. Krakow} {\bf 2}, 1 (1892).

\bibitem{1930Perron}
O.~{Perron}, {\em Mathematische Zeitschrift} {\bf {32, 5}}, 702 (1930).

\bibitem{2015ComAC...2....2B}
T.~{Boekholt}, S.~{Portegies Zwart}, {\em Computational Astrophysics and
  Cosmology} {\bf 2}, 2 (2015).

\bibitem{2014ApJ...785L...3P}
S.~{Portegies Zwart}, T.~{Boekholt}, {\em \apjl} {\bf 785}, L3 (2014).

\bibitem{Szebehely1991}
V.~Szebehely, {\em Chaos, Stability and Predictability in Newtonian Dynamics},
  pp. 63--71. Springer US, Boston, MA (1991).

\bibitem{2008MNRAS.388..965L}
H.~J. {Lehto}, et~al., {\em \mnras} {\bf 388}, 965 (2008).

\bibitem{2016ttp..book.....V}
M.~{Valtonen}, et~al., {\em {The Three-body Problem from Pythagoras to
  Hawking}} (2016).

\bibitem{1991pscn.proc...47H}
D.~C. {Heggie}, in S.~{Roeser}, U.~{Bastian} (eds.), {\em Predictability,
  Stability, and Chaos in N-Body Dynamical Systems}, pp. 47--62 (1991).

\bibitem{1994CeMDA..60..365A}
J.~P. {Anosova}, V.~V. {Orlov}, S.~J. {Aarseth}, {\em Celestial Mechanics and
  Dynamical Astronomy} {\bf 60}, 365 (1994).

\bibitem{LichtenbergAndLieberman1992}
M.~A. {Lichtenberg}, A.~{Lieberman}, {\em {Regular and Chaotic Dynamics}}.
  Springer (1992).

\bibitem{10.2307/24893288}
J.~Casasayas, J.~Llibre, {\em Indiana University Mathematics Journal} {\bf
  31}(4), 463 (1982).

\bibitem{1974InMat..27..191M}
R.~{McGehee}, {\em Inventiones Mathematicae} {\bf 27}, 191 (1974).

\bibitem{1913AN....195..113B}
C.~{Burrau}, {\em Astronomische Nachrichten} {\bf 195}, 113 (1913).

\bibitem{1967AJ.....72..876S}
V.~{Szebehely}, C.~F. {Peters}, {\em \aj} {\bf 72}, 876 (1967).

\bibitem{1994CeMDA..58....1A}
S.~J. {Aarseth}, J.~P. {Anosova}, V.~V. {Orlov}, V.~G. {Szebehely}, {\em
  Celestial Mechanics and Dynamical Astronomy} {\bf 58}, 1 (1994).

\bibitem{2016MNRAS.461.3576B}
T.~C.~N. {Boekholt}, F.~I. {Pelupessy}, D.~C. {Heggie}, S.~F. {Portegies
  Zwart}, {\em \mnras} {\bf 461}, 3576 (2016).

\bibitem{2013CoPhC.183..456P}
S.~{Portegies Zwart}, S.~L.~W. {McMillan}, E.~{van Elteren}, I.~{Pelupessy},
  N.~{de Vries}, {\em Computer Physics Communications} {\bf 183}, 456 (2013).

\bibitem{2013A&A...557A..84P}
F.~I. {Pelupessy}, et~al., {\em \aap} {\bf 557}, A84 (2013).

\bibitem{springerlink:10.1007/BF01386092}
R.~Bulirsch, J.~Stoer, {\em Numerische Mathematik} {\bf 6}, 413 (1964),
  10.1007/BF01386092.

\bibitem{1971Ap&SS..14..151H}
M.~H. {H{\'e}non}, {\em \apss} {\bf 14}, 151 (1971).

\bibitem{1983ApJ...268..319H}
P.~{Hut}, J.~N. {Bahcall}, {\em \apj} {\bf 268}, 319 (1983).

\bibitem{2008MNRAS.386..425F}
N.~T. {Faber}, C.~M. {Boily}, S.~{Portegies Zwart}, {\em \mnras} {\bf 386}, 425
  (2008).

\bibitem{hollinger2012nature}
H.~B. Hollinger, M.~Zenzen, {\em The nature of irreversibility: a study of its
  dynamics and physical origins}, vol.~28. Springer Science \& Business Media
  (2012).

\bibitem{MR103254}
A.~N. Kolmogorov, {\em Dokl. Akad. Nauk SSSR} {\bf 119}, 861 (1958).

\bibitem{MR103256}
J.~Sinai, {\em Dokl. Akad. Nauk SSSR} {\bf 124}, 768–771 (1959).

\bibitem{2017arXiv170407715R}
H.~{Rein}, D.~{Tamayo}, {\em ArXiv e-prints}  (2017).

\bibitem{1992PhyD...56....1E}
D.~J.~D. {Earn}, S.~{Tremaine}, {\em Physica D Nonlinear Phenomena} {\bf 56}, 1
  (1992).

\bibitem{1992PASJ...44..141M}
J.~{Makino}, S.~J. {Aarseth}, {\em \pasj} {\bf 44}, 141 (1992).

\bibitem{AMUSE}
S.~{Portegies Zwart}, S.~{McMillan}, {\em {Astrophysical Recipes: the Art of
  AMUSE}}. AAS IOP Astronomy (2017).

\bibitem{PhysRev.159.98}
L.~Verlet, {\em Phys. Rev.} {\bf 159}, 98 (1967).

\bibitem{1999A&A...348..117P}
S.~F. {Portegies Zwart}, J.~{Makino}, S.~L.~W. {McMillan}, P.~{Hut}, {\em \aap}
  {\bf 348}, 117 (1999).

\bibitem{1987PhLA..122..399P}
J.~I. {Palmore}, J.~L. {McCauley}, {\em Physics Letters A} {\bf 122}, 399
  (1987).

\bibitem{KlingonDictionary}
M.~Okrand, {\em {The Klingon Dictionary}}. The Klingon Language Institute
  (1992).

\bibitem{1992PhLA..166..340F}
S.~T. {Fryska}, M.~A. {Zohdy}, {\em Physics Letters A} {\bf 166}, 340 (1992).

\bibitem{2003PhRvL..90e4104H}
W.~B. {Hayes}, {\em Physical Review Letters} {\bf 90}(5), 054104 (2003).

\bibitem{2008DDA....39.1514V}
D.~{Veras}, E.~B. {Ford}, in {\em AAS/Division of Dynamical Astronomy Meeting
  \#39}, vol.~39 of {\em AAS/Division of Dynamical Astronomy Meeting}, p. 15.14
  (2008).

\bibitem{2015ApJ...798....8Z}
R.~E. {Zeebe}, {\em \apj} {\bf 798}, 8 (2015).

\bibitem{1986LNP...267..233H}
D.~C. {Heggie}, R.~D. {Mathieu}, in P.~{Hut}, S.~L.~W. {McMillan} (eds.), {\em
  The Use of Supercomputers in Stellar Dynamics}, vol. 267 of {\em Lecture
  Notes in Physics, Berlin Springer Verlag}, p. 233 (1986).

\end{thebibliography}
\end{document}